\documentclass[a4paper]{amsart}
\usepackage{epic,eepic,amsmath,amsthm,amssymb,epsfig,cite,indentfirst}
\usepackage{multicol,boxedminipage}

\begin{document}

\title[DW spin-$k/2$ vertex models]
{Higher spin vertex models with domain wall boundary conditions}

\author[Caradoc]{A Caradoc}
\address{Department of Mathematics and Statistics,
         University of Melbourne,
         Parkville, Victoria 3010, Australia.}
\email{a.caradoc@ms.unimelb.edu.au}

\author[Foda]{O Foda}

\address{Department of Mathematics and Statistics,
         University of Melbourne, 
	 Parkville, Victoria 3010, Australia.}
\email{foda@ms.unimelb.edu.au}

\author[Kitanine]{N Kitanine}

\address{LPTM (UCP--CNRS)
Universit\'e de Cergy-Pontoise,
Site de Saint-Martin, Pontoise,
95302 Cergy-Pontoise cedex}
\email{nikolai.kitanine@ptm.u-cergy.fr}

\date{}

\begin{abstract}

We derive determinant expressions for the partition functions of 
spin-$k/2$ vertex models on a finite square lattice with domain 
wall boundary conditions.

\end{abstract}

\maketitle

\newtheorem{theo}{Theorem}
\newtheorem{co}{Corollary}[theo]
\newtheorem{de}{Definition}
\newtheorem{pr}{Proposition}
\newtheorem{re}{Remark}
\newtheorem{ex}{Example}
\newtheorem{no}{Notation}
\newtheorem{ca}{Figure}

\def\ll{\left\lgroup}
\def\rr{\right\rgroup}

\newtheorem{definition}{Definition}
\newtheorem{theorem}{Theorem}
\newtheorem{remark}{Remark}
\newtheorem{lemma}{Lemma}
\newtheorem{proposition}{Proposition}
\newtheorem{corollary}{Corollary}
\newtheorem{example}{Example}

\newcommand{\field}[1]{\mathbb{#1}}

\newcommand{\Proof}{\medskip\noindent {\it Proof: }}

\def\beqa{\begin{eqnarray}}
\def\eeqa{\end{eqnarray}}
\def\ba{\begin{array}}
\def\ea{\end{array}}
\def\r{\rangle}
\def\l{\langle}
\def\a{\alpha}
\def\b{\beta}
\def\hb{\hat\beta}
\def\d{\delta}
\def\g{\gamma}
\def\e{\epsilon}
\def\tg{\operatorname{tg}}
\def\ctg{\operatorname{ctg}}
\def\sh{\operatorname{sh}}
\def\ch{\operatorname{ch}}
\def\cth{\operatorname{cth}}
\def\th{\operatorname{th}}
\def\eps{\varepsilon}
\def\la{\lambda}
\def\tla{\tilde{\lambda}}
\def\tmu{\tilde{\mu}}
\def\s{\sigma}
\def\sul{\sum\limits}
\def\pl{\prod\limits}
\def\lt({\left(}
\def\rt){\right)}
\def\pd #1{\frac{\partial}{\partial #1}}
\def\const{{\rm const}}
\def\argum{\{\mu_j\},\{\la_k\}} 
\def\umarg{\{\la_k\},\{\mu_j\}} 
\def\prodmu #1{\prod\limits_{j #1 k} \sinh(\mu_k-\mu_j)}
\def\prodla #1{\prod\limits_{j #1 k} \sinh(\lambda_k-\lambda_j)}
\newcommand{\bl}[1]{\makebox[#1em]{}}
\def\tr{\operatorname{tr}}
\def\Res{\operatorname{Res}}
\def\det{\operatorname{det}}

\newcommand{\boldN}{\boldsymbol{N}}
\newcommand{\bra}[1]{\langle\,#1\,|}
\newcommand{\ket}[1]{|\,#1\,\rangle}
\newcommand{\bracket}[1]{\langle\,#1\,\rangle}
\newcommand{\infinity}{\infty}

\renewcommand{\labelenumi}{\S\theenumi.}

\let\up=\uparrow
\let\down=\downarrow
\let\tend=\rightarrow
\hyphenation{boson-ic
             ferm-ion-ic
             para-ferm-ion-ic
             two-dim-ension-al
             two-dim-ension-al
	     rep-resent-ative}

\setcounter{section}{-1}

\section{Introduction}\label{introduction}

In \cite{korepin}, Korepin introduced the concept of domain wall 
(DW) boundary conditions for the six vertex (or spin-$1/2$) model 
on a finite lattice, and proposed recursion relations that fully 
determine the partition function in that case. In \cite{izergin}, 
Izergin obtained a determinant solution of Korepin's recursion 
relations. In this work, inspired by Slavnov's inner product 
formula for higher spin models \cite{slavnov}, we derive 
determinant expressions for the spin-$k/2$ DW partition 
functions\footnote{In the sequel, {\it partition function} 
will refer to {\it DW partition function} unless otherwise 
indicated. For applications of the Izergin-Korepin determinant 
formula in statistical mechanics, see \cite{korepin-book}. For 
applications to algebraic combinatorics, see \cite{bressoud-book}.}, 
$k \in \boldN$, using the fact that these models are related to the 
spin-$1/2$ model using fusion \cite{rks, djkmo}.

Basically, we show that fusion can be applied at the level of 
spin-$1/2$ partition functions to obtain spin-$k/2$ partition 
functions, for any $k \in \boldN$. More specifically, our result, 
in words, is that appropriate specializations of the rapidity 
variables in Izergin's determinant expression for the spin-$1/2$ 
partition function on a $kL\times kL$ lattice (followed by suitable 
normalizations) yield determinant expressions for the spin-$k/2$ 
partition functions on an $L\times L$ lattice.

In sections \ref{vertex-models} and 
\ref{fusion}, we briefly introduce the spin-$k/2$ vertex models, 
and outline the fusion procedure. In \ref{motivation} and 
\ref{proof}, we motivate our result, then outline its proof. 
In \ref{example}, we give the spin-1 partition function 
as a specific example of our general result, that has the 
benefit of allowing for an independent Izergin-type proof 
(which is not available for higher spin models).
In \ref{hom-lim}, we derive the homogeneous limit of the
spin-$k/2$ result, and comment on combinatorics of higher 
spin models. Finally, an appendix contains technical details. 
The presentation is elementary and (almost) self-contained.

\section{Vertex models}\label{vertex-models}

\subsection*{Definitions related to vertex models} 

We work on a square lattice consisting of $L$ horizontal lines 
(labelled from bottom to top), $L$ vertical lines (labelled 
from left to right) and $L^2$ intersection points.

We assign the {\it i}-th horizontal line an orientation from left 
to right, and a complex rapidity variable $x_i$. We assign the 
{\it j}-th vertical line an orientation from bottom to top, and 
a complex rapidity variable $y_j$. All rapidity variables are 
independent, unless specifically indicated to be otherwise. 

\noindent $\{{\bf x}\}$ is a set of rapidity variables 
$\{x_1, \cdots, x_C\}$, where the cardinality of the set should 
be clear from context. 

\noindent $\{{\bf x}\}_i$ is a set $\{x_1, x_2, \cdots, x_C\}$, 
but with the $x_i$ element missing. 

\noindent {\bf A $k$-stack $\{{\bf x}| k\}$} is a set of 
$k$ variables of the form $\{x, x+1, x+2, \cdots, x+k-1\}$.

\begin{multicols}{2}

\noindent{\bf A bond} is a line segment between two intersection 
points. 

\noindent{\bf A boundary, or extremal bond} is a line segment, at 
the boundary of a line, attached to a single intersection point. 

In a spin-$k/2$ models, we assign each bond $\kappa$ arrows, where 
$\kappa$ $\in$ $\{k, k-2, \cdots, k \mod 2\}$. 
All $\kappa$ arrows, on the same bond, point in the same direction.

The $\kappa$ arrows on a bond define a {\it spin} variable on that 
bond. 

\noindent{\bf The magnitude of spin} on a bond is ${\kappa}/{2}$. 

\noindent{\bf The sign of spin} on a bond is positive (negative) 
if the $\kappa$ arrows are oriented in the same (opposite) direction 
as (to) the rapidity flow in that bond. 

%
\begin{center}
\begin{minipage}{2in}
\setlength{\unitlength}{0.001cm}
\begin{picture}(4800, 6000)(0, 0)
\thicklines
\path(2400,5400)(2400,1800)
\path(3000,5400)(3000,1800)
\path(3600,5400)(3600,1800)
\path(4200,5400)(4200,1800)
\path(4800,5400)(4800,1800)
\path(1800,4800)(5400,4800)
\path(1800,4200)(5400,4200)
\path(1800,3600)(5400,3600)
\path(1800,3000)(5400,3000)
\path(1800,2400)(5400,2400)
\path(0600,4254)(1200,4254)
\path(2400,654)(2400,1254)
\path(3000,654)(3000,1254)
\path(3600,654)(3600,1254)
\path(4200,654)(4200,1254)
\path(4800,654)(4800,1254)
\path(600,2454)(1200,2454)
\path(600,3054)(1200,3054)
\path(600,3654)(1200,3654)
\path(600,4854)(1200,4854)
\whiten\path(2490,894)(2400,1254)(2310,894)(2490,894)
\whiten\path(3090,894)(3000,1254)(2910,894)(3090,894)
\whiten\path(3690,894)(3600,1254)(3510,894)(3690,894)
\whiten\path(4290,894)(4200,1254)(4110,894)(4290,894)
\whiten\path(4890,894)(4800,1254)(4710,894)(4890,894)
\whiten\path(840,2364)(1200,2454)(840,2544)(840,2364)
\whiten\path(840,2964)(1200,3054)(840,3144)(840,2964)
\whiten\path(840,3564)(1200,3654)(840,3744)(840,3564)
\whiten\path(840,4164)(1200,4254)(840,4344)(840,4164)
\whiten\path(840,4764)(1200,4854)(840,4944)(840,4764)
\put(0100,4854){$x_L$}
\put(0100,3054){$x_2$}
\put(0100,2454){$x_1$}
\put(2300,0250){$y_1$}
\put(2900,0250){$y_2$}
\put(4700,0250){$y_L$}
\end{picture}
\begin{ca} 
\label{lattice}
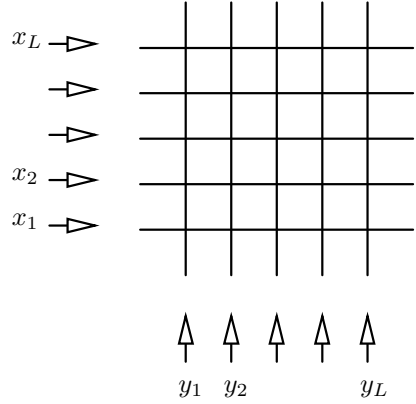
A finite square lattice, with oriented lines and 
rapidities. 
\end{ca}
\end{minipage}
\end{center}

\end{multicols}

\noindent{\bf A vertex} $v_{ij}$, is the intersection point of 
the $i$-th horizontal line and the $j$-th vertical line, together 
with the 4 bonds attached to it, and the arrows on them. 

\noindent {\bf a weight}, $w_{ij}$, is a function assigned to 
a vertex, $v_{ij}$, that depends on the difference of rapidity 
variables flowing through that vertex. In exactly solvable models, 
the weights satisfy the Yang Baxter equations \cite{baxter-book}.

\subsection*{Frequently used abbreviations}
We frequently use the bracket notation ${\bf [x]}= \sinh(\lambda x)$ 
(where $\lambda$ is a constant, {\it `crossing'} parameter that 
characterizes the model) and the related product notation
${\bf [x]}_{m} = [x][x-1] \cdots [x-m+1]$. We also use the 
abbreviations ${\tilde{x}} = x+1$ and $u_{ij}=-x_i + y_j$.

\noindent{\bf The partition function} of a spin-$k/2$ vertex model, 
on an $L\times L$ lattice, $Z^{k/2\times k/2}_{L\times L}$, is 
a weighted sum over all configurations, that satisfy certain 
boundary conditions. The weight of a configuration is the product 
of the weights, $w_{ij}$, of the vertices $v_{ij}$. 

\begin{equation}
\label{partitionfunctionequation}
Z^{k/2\times k/2}_{L\times L}\ll \{{\bf x}\}, \{{\bf y}\}\rr = 
\sum
\pl_{\rm vertices} w_{ij}
\end{equation}

\subsection*{Conservation of spin flow} The vertex models, that we 
discuss in this work, conserve spin flow: In all vertices that have 
non-zero weight, the net incoming spin (measured with respect to 
rapidity inflow) equals the net outgoing spin (measured with respect 
to rapidity outflow).

\subsection*{The spin-$1/2$ model} There are six vertex types that 
conserve spin flow in the spin-$1/2$ model. They are shown in figure 
\ref{sixvertices} below. For convenience, we label the vertices by 
their weights: an $a$ vertex has weight $a(x, y)$, and so forth. 

We do not need to distinguish vertices that share the same weight, 
except in one case (the $c+$ vertex) mentioned below. The six vertices 
of the spin-$1/2$ and their weights are shown in figure \ref{sixvertices}. 
The weights of every two vertices in the same column are equal and shown 
below them.


\begin{center}
\begin{minipage}{4in}

\setlength{\unitlength}{0.0008cm}
\begin{picture}(8000,7500)(0, 0)
\thicklines

\blacken\path(10162,2310)(10522,2400)(10162,2490)(10162,2310)
\blacken\path(1282,2490)(0922,2400)(1282,2310)(1282,2490)
\blacken\path(1462,5610)(1822,5700)(1462,5790)(1462,5610)
\blacken\path(1732,1860)(1822,1500)(1912,1860)(1732,1860)
\blacken\path(1732,2685)(1822,2325)(1912,2685)(1732,2685)
\blacken\path(1912,5340)(1822,5700)(1732,5340)(1912,5340)
\blacken\path(1912,6240)(1822,6600)(1732,6240)(1912,6240)
\blacken\path(2182,2490)(1822,2400)(2182,2310)(2182,2490)
\blacken\path(2362,5610)(2722,5700)(2362,5790)(2362,5610)
\blacken\path(5182,5790)(4822,5700)(5182,5610)(5182,5790)
\blacken\path(5362,2310)(5722,2400)(5362,2490)(5362,2310)
\blacken\path(5632,1860)(5722,1500)(5812,1860)(5632,1860)
\blacken\path(5632,2760)(5722,2400)(5812,2760)(5632,2760)
\blacken\path(5812,5340)(5722,5700)(5632,5340)(5812,5340)
\blacken\path(5812,6240)(5722,6600)(5632,6240)(5812,6240)
\blacken\path(6082,5790)(5722,5700)(6082,5610)(6082,5790)
\blacken\path(6262,2310)(6622,2400)(6262,2490)(6262,2310)
\blacken\path(9082,2490)(8722,2400)(9082,2310)(9082,2490)
\blacken\path(9262,5610)(9622,5700)(9262,5790)(9262,5610)
\blacken\path(9532,2760)(9622,2400)(9712,2760)(9532,2760)
\blacken\path(9532,5160)(9622,4800)(9712,5160)(9532,5160)
\blacken\path(9712,2040)(9622,2400)(9532,2040)(9712,2040)
\blacken\path(9712,6240)(9622,6600)(9532,6240)(9712,6240)
\blacken\path(9982,5790)(9622,5700)(9982,5610)(9982,5790)
\path(10222,5700)(9622,5700)
\path(1222,5700)(1822,5700)
\path(1522,2400)(0922,2400)
\path(1822,1500)(1822,2100)
\path(1822,2325)(1822,2925)
\path(1822,3300)(1822,1500)
\path(1822,3900)(1822,4500)
\path(1822,5700)(1822,5100)
\path(1822,0600)(1822,1200)
\path(1822,6600)(1822,4800)
\path(0022,2400)(0622,2400)
\path(0022,5700)(0622,5700)
\path(2422,2400)(1822,2400)
\path(3922,2400)(4522,2400)
\path(3922,5700)(4522,5700)
\path(4822,2400)(6622,2400)
\path(4822,5700)(6622,5700)
\path(5122,2400)(5722,2400)
\path(5422,5700)(4822,5700)
\path(5722,1500)(5722,2100)
\path(5722,2400)(5722,3000)
\path(5722,3300)(5722,1500)
\path(5722,3900)(5722,4500)
\path(5722,5700)(5722,5100)
\path(5722,600)(5722,1200)
\path(5722,6600)(5722,4800)
\path(5722,6600)(5722,6000)
\path(6022,2400)(6622,2400)
\path(6322,5700)(5722,5700)
\path(7822,2400)(8422,2400)
\path(7822,5700)(8422,5700)
\path(8722,2400)(10522,2400)
\path(8722,5700)(10522,5700)
\path(9022,5700)(9622,5700)
\path(0922,2400)(2722,2400)
\path(0922,5700)(2722,5700)
\path(9322,2400)(8722,2400)
\path(9622,2400)(9622,1800)
\path(9622,2400)(9622,3000)
\path(9622,3300)(9622,1500)
\path(9622,3900)(9622,4500)
\path(9622,4800)(9622,5400)
\path(9622,0600)(9622,1200)
\path(9622,6600)(9622,4800)
\path(9622,6600)(9622,6000)
\path(9922,2400)(10522,2400)

\whiten\path(1912,4140)(1822,4500)(1732,4140)(1912,4140)
\whiten\path(1912,0840)(1822,1200)(1732,0840)(1912,0840)
\whiten\path(0262,2310)(0622,2400)(0262,2490)(0262,2310)
\whiten\path(0262,5610)(0622,5700)(0262,5790)(0262,5610)
\whiten\path(4162,2310)(4522,2400)(4162,2490)(4162,2310)
\whiten\path(4162,5610)(4522,5700)(4162,5790)(4162,5610)
\whiten\path(5812,4140)(5722,4500)(5632,4140)(5812,4140)
\whiten\path(5812,0840)(5722,1200)(5632,0840)(5812,0840)
\whiten\path(8062,2310)(8422,2400)(8062,2490)(8062,2310)
\whiten\path(8062,5610)(8422,5700)(8062,5790)(8062,5610)
\whiten\path(9712,4140)(9622,4500)(9532,4140)(9712,4140)
\whiten\path(9712,0840)(9622,1200)(9532,0840)(9712,0840)

\put(0000,0000){$a(x, y) = [u+1]$} 

\put(4500,0000){$b(x, y) = [u]$} 

\put(8500,0000){$c(x, y) = [1]$} 

\put(0300,5000){$x$}
\put(4200,5000){$x$}
\put(8100,5000){$x$}

\put(0300,1700){$x$}
\put(4200,1700){$x$}
\put(8100,1700){$x$}

\put(02000,4200){$y$}
\put(06000,4200){$y$}
\put(10000,4200){$y$}

\put(2000,0900){$y$}
\put(6000,0900){$y$}
\put(10000,0900){$y$}

\end{picture}

\begin{ca}
\label{sixvertices}
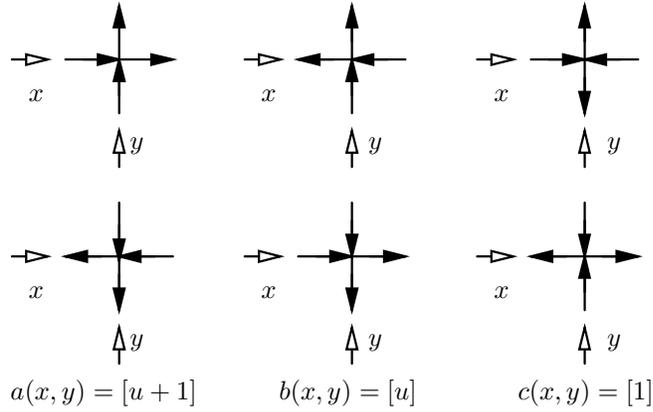
The six vertices of the spin-$1/2$ model and their weights. 
$u=-x+y$.
\end{ca}

\end{minipage}
\end{center}

\begin{multicols}{2}

\subsection*{The $c+$ vertex} The vertex with all arrows 
pointing inwards from left and right, and all arrows 
pointing outwards from above and below, plays a rather 
special role in this work. We refer to it as the $c+$ 
vertex. There is a unique $c+$ vertex in every spin-$k/2$
model. The spin-$2$ $c+$ vertex is shown in figure
\ref{spin-2-c+vertex}.

%
\begin{center}
\begin{minipage}{2in}
\setlength{\unitlength}{0.0008cm}
\begin{picture}(4000, 4000)(0, 0)
\thicklines
\path(0020,2220)(4020,2220)
\path(2020,0420)(2020,3820)
\blacken\path(0560,2130)(0920,2220)(0560,2310)(0560,2130)
\blacken\path(0920,2130)(1280,2220)(0920,2310)(0920,2130)
\blacken\path(1280,2130)(1640,2220)(1280,2310)(1280,2130)
\blacken\path(1640,2130)(2000,2220)(1640,2310)(1640,2130)
\blacken\path(2380,2310)(2020,2220)(2380,2130)(2380,2310)
\blacken\path(2740,2310)(2380,2220)(2740,2130)(2740,2310)
\blacken\path(3100,2310)(2740,2220)(3100,2130)(3100,2310)
\blacken\path(3460,2310)(3100,2220)(3460,2130)(3460,2310)
%
%
\blacken\path(2110,2540)(2020,2900)(1930,2540)(2110,2540)
\blacken\path(2110,2900)(2020,3260)(1930,2900)(2110,2900)
\blacken\path(2110,3260)(2020,3620)(1930,3260)(2110,3260)
\blacken\path(2110,3620)(2020,3980)(1930,3620)(2110,3620)
%
%
\blacken\path(1930,0780)(2020,0320)(2110,0780)(1930,0780)
\blacken\path(1930,1140)(2020,0780)(2110,1140)(1930,1140)
\blacken\path(1930,1500)(2020,1140)(2110,1500)(1930,1500)
\blacken\path(1930,1860)(2020,1500)(2110,1860)(1930,1860)
\end{picture}
\begin{ca}
\label{spin-2-c+vertex}
The spin-2 $c+$ vertex.
\end{ca}
\end{minipage}
\end{center}

\end{multicols}

\begin{multicols}{2}
\subsection*{Domain wall (DW) boundary conditions} Consider 
the spin-$1/2$ model on a finite square lattice, and require 
that the boundary arrows have the same orientation as the
arrow on the corresponding boundary of the $c+$ vertex: all 
arrows on the left and right boundaries point inwards, and 
all arrows on the upper and lower boundaries point outwards. 
The $c+$ vertex is a DW configuration on a $1\times 1$ lattice. 


\begin{center}
\begin{minipage}{2in}

\setlength{\unitlength}{0.0008cm}

\begin{picture}(4000,6000)(-200, 0)
\thicklines

\path(300,4800)(5700,4800)
\path(300,3900)(5700,3900)
\path(300,2100)(5700,2100)
\path(300,1200)(5700,1200)
\path(1200,5700)(1200,300)
\path(2100,5700)(2100,300)
\path(4800,5700)(4800,300)
\path(1200,5025)(1200,5625)

\blacken\path(1110,1560)(1200,1200)(1290,1560)(1110,1560)
\blacken\path(1110,2460)(1200,2100)(1290,2460)(1110,2460)
\blacken\path(1110,3360)(1200,3000)(1290,3360)(1110,3360)
\blacken\path(1110,0735)(1200,0375)(1290,0735)(1110,0735)
\blacken\path(1290,4515)(1200,4875)(1110,4515)(1290,4515)
\blacken\path(1290,5265)(1200,5625)(1110,5265)(1290,5265)
\blacken\path(1560,3990)(1200,3900)(1560,3810)(1560,3990)
\blacken\path(1665,2910)(2025,3000)(1665,3090)(1665,2910)
\blacken\path(1665,4710)(2025,4800)(1665,4890)(1665,4710)
\blacken\path(1740,1110)(2100,1200)(1740,1290)(1740,1110)
\blacken\path(1740,2010)(2100,2100)(1740,2190)(1740,2010)
\blacken\path(2010,1560)(2100,1200)(2190,1560)(2010,1560)
\blacken\path(2010,2460)(2100,2100)(2190,2460)(2010,2460)
\blacken\path(2010,0660)(2100,0300)(2190,0660)(2010,0660)
\blacken\path(2190,3615)(2100,3975)(2010,3615)(2190,3615)
\blacken\path(2190,4440)(2100,4800)(2010,4440)(2190,4440)
\blacken\path(2190,5265)(2100,5625)(2010,5265)(2190,5265)
\blacken\path(2385,3990)(2025,3900)(2385,3810)(2385,3990)
\blacken\path(2460,3090)(2100,3000)(2460,2910)(2460,3090)
\blacken\path(2640,1110)(3000,1200)(2640,1290)(2640,1110)
\blacken\path(2640,2010)(3000,2100)(2640,2190)(2640,2010)
\blacken\path(2640,4710)(3000,4800)(2640,4890)(2640,4710)
\blacken\path(2910,4260)(3000,3900)(3090,4260)(2910,4260)
\blacken\path(2910,0660)(3000,0300)(3090,0660)(2910,0660)
\blacken\path(3090,1740)(3000,2100)(2910,1740)(3090,1740)
\blacken\path(3090,2640)(3000,3000)(2910,2640)(3090,2640)
\blacken\path(3090,3540)(3000,3900)(2910,3540)(3090,3540)
\blacken\path(3090,5340)(3000,5700)(2910,5340)(3090,5340)
\blacken\path(3360,1290)(3000,1200)(3360,1110)(3360,1290)
\blacken\path(3360,3090)(3000,3000)(3360,2910)(3360,3090)
\blacken\path(3435,4890)(3075,4800)(3435,4710)(3435,4890)
\blacken\path(3540,2010)(3900,2100)(3540,2190)(3540,2010)
\blacken\path(3540,3810)(3900,3900)(3540,3990)(3540,3810)
\blacken\path(3810,1560)(3900,1200)(3990,1560)(3810,1560)
\blacken\path(3810,2460)(3900,2100)(3990,2460)(3810,2460)
\blacken\path(3810,3360)(3900,3000)(3990,3360)(3810,3360)
\blacken\path(3810,0660)(3900,0300)(3990,0660)(3810,0660)
\blacken\path(3990,4440)(3900,4800)(3810,4440)(3990,4440)
\blacken\path(3990,5265)(3900,5625)(3810,5265)(3990,5265)
\blacken\path(4260,3990)(3900,3900)(4260,3810)(4260,3990)
\blacken\path(4260,4890)(3900,4800)(4260,4710)(4260,4890)
\blacken\path(4335,1290)(3975,1200)(4335,1110)(4335,1290)
\blacken\path(4335,3090)(3975,3000)(4335,2910)(4335,3090)
\blacken\path(4440,2010)(4800,2100)(4440,2190)(4440,2010)
\blacken\path(4710,1560)(4800,1200)(4890,1560)(4710,1560)
\blacken\path(4710,0660)(4800,0300)(4890,0660)(4710,0660)
\blacken\path(4890,2640)(4800,3000)(4710,2640)(4890,2640)
\blacken\path(4890,3615)(4800,3975)(4710,3615)(4890,3615)
\blacken\path(4890,4440)(4800,4800)(4710,4440)(4890,4440)
\blacken\path(4905,5340)(4800,5700)(4695,5340)(4905,5340)
\blacken\path(5160,2190)(4800,2100)(5160,2010)(5160,2190)
\blacken\path(5160,3990)(4800,3900)(5160,3810)(5160,3990)
\blacken\path(5160,4890)(4800,4800)(5160,4710)(5160,4890)
\blacken\path(5235,1290)(4875,1200)(5235,1110)(5235,1290)
\blacken\path(5235,3090)(4875,3000)(5235,2910)(5235,3090)
\blacken\path(0765,3810)(1125,3900)(0765,3990)(0765,3810)
\blacken\path(0765,4710)(1125,4800)(0765,4890)(0765,4710)
\blacken\path(0840,1110)(1200,1200)(0840,1290)(0840,1110)
\blacken\path(0840,2010)(1200,2100)(0840,2190)(0840,2010)
\blacken\path(0840,2910)(1200,3000)(0840,3090)(0840,2910)

\path(1200,1800)(1200,1200)
\path(1200,2700)(1200,2100)
\path(1200,3600)(1200,3000)
\path(1200,4275)(1200,4875)
\path(1200,0975)(1200,0375)
\path(1425,3000)(2025,3000)
\path(1425,4800)(2025,4800)
\path(1500,1200)(2100,1200)
\path(1500,2100)(2100,2100)
\path(1800,3900)(1200,3900)
\path(2100,1800)(2100,1200)
\path(2100,2700)(2100,2100)
\path(2100,3375)(2100,3975)
\path(2100,4200)(2100,4800)
\path(2100,5025)(2100,5625)
\path(2100,0900)(2100,0300)
\path(2400,1200)(3000,1200)
\path(2400,2100)(3000,2100)
\path(2400,4800)(3000,4800)
\path(2625,3900)(2025,3900)
\path(2700,3000)(2100,3000)
\path(3000,1500)(3000,2100)
\path(3000,2400)(3000,3000)
\path(3000,3300)(3000,3900)
\path(3000,4500)(3000,3900)
\path(3000,5100)(3000,5700)
\path(3000,5700)(3000,0300)
\path(3000,0900)(3000,0300)
\path(0300,3000)(5700,3000)
\path(3300,2100)(3900,2100)
\path(3300,3900)(3900,3900)
\path(3600,1200)(3000,1200)
\path(3600,3000)(3000,3000)
\path(3675,4800)(3075,4800)
\path(3900,1800)(3900,1200)
\path(3900,2700)(3900,2100)
\path(3900,3600)(3900,3000)
\path(3900,4200)(3900,4800)
\path(3900,5025)(3900,5625)
\path(3900,5700)(3900,0300)
\path(3900,0900)(3900,0300)
\path(4200,2100)(4800,2100)
\path(4500,3900)(3900,3900)
\path(4500,4800)(3900,4800)
\path(4575,1200)(3975,1200)
\path(4575,3000)(3975,3000)
\path(4800,1800)(4800,1200)
\path(4800,2400)(4800,3000)
\path(4800,3375)(4800,3975)
\path(4800,4200)(4800,4800)
\path(4800,5100)(4800,5700)
\path(4800,0900)(4800,0300)
\path(0525,3900)(1125,3900)
\path(0525,4800)(1125,4800)
\path(5400,2100)(4800,2100)
\path(5400,3900)(4800,3900)
\path(5400,4800)(4800,4800)
\path(5475,1200)(4875,1200)
\path(5475,3000)(4875,3000)
\path(0600,1200)(1200,1200)
\path(0600,2100)(1200,2100)
\path(0600,3000)(1200,3000)

%
%
%

\end{picture}

\begin{ca}
\label{dwbc}
A DW configuration.
\end{ca}

\end{minipage}
\end{center}
\end{multicols}

\noindent{\bf The Izergin-Korepin (IK) determinant expression} 
for the spin-$1/2$ partition function \cite{korepin,izergin} is

\begin{equation}
Z^{1/2\times 1/2}_{L\times L}\ll \{{\bf x}\}, \{{\bf y}\} \rr 
=
\frac{\pl_{{i, j}=1}^{L} [- x_i + y_j + 1]_{2}}
     {\pl_{1 \le i < j \le L}[- x_i + x_j][- y_j + y_i]} 
      det \ll M^{1/2\times 1/2}_{L\times L}\rr
\end{equation}
\noindent where

$$
M^{1/2\times 1/2}_{L\times L, ij} = \frac{[1]}{[- x_{i}+ y_{j}+1]_{2}}
$$

\subsection*{Definitions related to partition functions, matrices and 
normalizations}

\noindent{\bf $Z^{k/2\times k/2}_{L\times L}$} is the partition 
function of the spin-$k/2$ model on an $L\times L$ lattice.

\noindent{\bf $M^{k/2\times k/2}_{L\times L}$} is the $L\times L$ IK
matrix, but with a $k\times k$ block structure, as will be explained 
below. 

\noindent{\bf $M^{k/2\times k/2}_{kL\times kL, ij}$} is the $k\times k$
$ij$-th block of $M^{k/2\times k/2}_{kL\times kL}$. It depends on the 
rapidities $\{x_i, y_j\}$.

\noindent{\bf $B^{k\times k}_{ij}$} is the $ij$-th $k\times k$ block 
of the lattice. 

\noindent{\bf $N^{k/2\times k/2}_{kL\times kL}$} is the normalization
function of $M^{k/2\times k/2}_{L\times L}$ that sets the weights 
of the spin-$k/2$ vertices to $[k]_k$.

\subsection*{Spin-$k/2$ models}\label{spin-k/2-models}

We will not need the weights of spin-$k/2$ model in all 
generality. They can be deduced from those of the fused 
elliptic height models, \cite{djkmo}, as follows.

\subsection*{Spin-$k/2$ vertex weights from elliptic height
weights}

\begin{enumerate}

\item Take the trigonometric limit: set the elliptic nome, which 
appears in the weights of elliptic models, $q$ $\rightarrow 0$. 
This reduces the weights to ratios of products of trigonometric 
functions (for pure imaginary values of the crossing parameter).

\item Take the vertex limit: set the height shift parameter, 
that appears in the weights of height models, $\zeta$ $\rightarrow 
\pm \infty$. This eliminates dependence on the height variables.

\item Symmetrize the resulting weights so that the weights 
of the $c$-type vertices are equal.

\end{enumerate}

\section{Fusion}\label{fusion}

As the spin-$k/2$ vertices are obtained from the spin-$1/2$ 
vertices using fusion, we wish to recall how fusion works, 
following \cite{djkmo}\footnote{Although \cite{djkmo} studies 
fusion in the context of elliptic height models, it is convenient 
to start from there, as their exposition is explicit.}.

\subsection*{Definitions related to boundaries} 

\noindent{\bf A boundary} of length $L$ is a set of $L$ 
parallel extremal bonds. A vertical boundary consists of 
horizontal extremal bonds. A horizontal boundary consists 
of vertical extremal bonds. Notice that, in our definition, 
a boundary cannot be closed. A closed boundary will consist 
of a sequence of horizontal and vertical boundaries.

\noindent{\bf A $\sigma$-configuration} is an arrangement 
of spins on a boundary, with total spin $\sigma$. A boundary 
of length $L$, in a spin-$k/2$ model, has 
$\sigma$ $\in$ $\{kL/2, kL/2 -1, \cdots, - kL/2\}$.

\noindent{\bf A $\sigma$-set} is the set of all 
$\sigma$-configurations on a boundary.

\noindent{\bf A $\sigma$-representative} of a $\sigma$-set, is any 
single uniquely defined configuration in that set, that we select 
to represent the entire set. In this work, we choose that to be the 
configuration with spins ordered in terms of their values, with larger 
(more positive) spins to the left of (lower than) smaller (less positive) 
spins in the case of horizontal (vertical) boundaries. 

\begin{multicols}{2}

\noindent{\bf An inflow (outflow) boundary} is one that rapidity variables 
flow into (out of), as seen from the inside of the region that it bounds.


\begin{center}
\begin{minipage}{2in}

\setlength{\unitlength}{0.0008cm}

\begin{picture}(3000, 2800)(0, 0)
\Thicklines

\path(0900,1500)(2700,1500)
\path(1800,2400)(1800,0600)

\put(0000,1500){$\alpha/2$}
\put(3100,1500){$\gamma/2$}

\put(1300,2900){$\delta/2$}
\put(1300,0000){$\beta/2 $}

\end{picture}

\begin{ca}
\label{general-vertex}
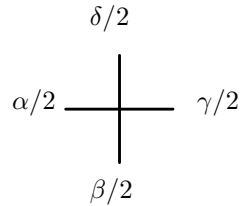
Spin values of a general vertex in a spin-$k/2$ model.
\end{ca}

\end{minipage}
\end{center}

\end{multicols}

\subsection*{Fusion procedure}

Following \cite{djkmo}, to compute the weight of the generic spin-$k/2$ 
vertex shown in figure \ref{general-vertex}, where 
$(\alpha, \beta, \gamma, \delta)$ 
$\in$ $\{k, k-1, \cdots, -k\}$ and $\alpha + \beta = \gamma + \delta$, 
we start from the set of all spin-$1/2$ configurations on a $k\times k$ 
lattice, with boundaries that match in total spin values those of the 
vertex that we wish to produce (the right boundary has total spin 
$\alpha/2$, {\it etc}), and proceed as follows.

\begin{enumerate}

\item Set $\{x_1,$ $ x_2,$ $     \cdots,$ $ x_k        \}$ 
      to $\{{\bf x_1} | k\}$, 
      and set 
          $\{y_1,$ $ y_2,$ $     \cdots,$ $ y_k        \}$ 
      to $\{{\bf y_1} | k\}$.
      $x_1$ and $y_1$ will be the rapidities of the resulting 
      spin-$k/2$ vertex. 

\item Sum over all $\alpha/2$-configurations on the left 
     (inflow) boundary. 

\item Sum over all $\beta/2$-configurations on the lower 
     (inflow) boundary. 

\item Take the $\gamma/2$-configuration on the right (outflow)
      boundary to be the unique $\gamma/2$-representative. 
      No summation over configurations is performed.

\item Take the $\delta/2$-configuration on the upper (outflow) 
      boundary to be the unique $\delta/2$-representative. 
      No summation over configurations is performed.

\item Normalize the result\footnote{Our normalization is different 
from that of \cite{djkmo}. We choose to normalize the $c+$ vertex 
of the spin-$k/2$ model to $[k]_k$. In \cite{djkmo}, the $c+$ vertices 
are normalized to 1 (up to phases), in the trigonometric vertex limit 
that we are interested in.}, by dividing with 

\begin{equation}\label{firstnormalization}
N^{k/2\times k/2}_{1\times 1}(x_1, y_1) = 
    {\pl_{p = 0}^{k-1}[-x_1 + y_1 + p]_{k-1}}
\end{equation}

\end{enumerate}


\subsection*{Remarks} To perform fusion, in our convention, inflow 
boundaries are summed over, while outflow boundaries are set to 
representative configurations. Further, we could have set the 
outflow boundaries to any $\sigma$-configuration with the correct 
net spin $\sigma$. However, using the Yang-Baxter equations, one 
can show that the result is independent of the choice \cite{djkmo}. 

\section{Spin-$k/2$ partition functions: Motivation}\label{motivation}

\begin{multicols}{2}

Suppose we wish to obtain the spin-2 $c+$ vertex of figure 
\ref{spin-2-c+vertex}. Following the fusion procedure, we 
need to consider the $4\times 4$ DW spin-$1/2$ partition 
function, shown in figure \ref{4-by-4-spin-1/2}. 

But this case is very simple: Because of the boundary 
conditions, all $\sigma$-sets have exactly one configuration
each, and we do not need to sum over left and lower 
configurations, or choose any right and upper ones.
All we need to do is to set the rapidities to the right 
values, and normalize suitably. 


\begin{center}
\begin{minipage}{2in}

\setlength{\unitlength}{0.0008cm}

\begin{picture}(3000, 5500)(1000, 500)
\thicklines


\blacken\path(1665,4710)(2025,4800)(1665,4890)(1665,4710)
\blacken\path(1665,3810)(2025,3900)(1665,3990)(1665,3810)
\blacken\path(1665,2910)(2025,3000)(1665,3090)(1665,2910)
\blacken\path(1665,2010)(2025,2100)(1665,2190)(1665,2010)


\blacken\path(2010,1560)(2100,1200)(2190,1560)(2010,1560)
\blacken\path(2910,1560)(3000,1200)(3090,1560)(2910,1560)
\blacken\path(3810,1560)(3900,1200)(3990,1560)(3810,1560)
\blacken\path(4710,1560)(4800,1200)(4890,1560)(4710,1560)


\blacken\path(4905,5340)(4800,5700)(4695,5340)(4905,5340)
\blacken\path(4005,5340)(3900,5700)(3795,5340)(4005,5340)
\blacken\path(3105,5340)(3000,5700)(2895,5340)(3105,5340)
\blacken\path(2205,5340)(2100,5700)(1995,5340)(2205,5340)


\blacken\path(5160,2190)(4800,2100)(5160,2010)(5160,2190)
\blacken\path(5160,3090)(4800,3000)(5160,2910)(5160,3090)
\blacken\path(5160,3990)(4800,3900)(5160,3810)(5160,3990)
\blacken\path(5160,4890)(4800,4800)(5160,4710)(5160,4890)


\path(1200,2100)(5700,2100)
\path(1200,3000)(5700,3000)
\path(1200,3900)(5700,3900)
\path(1200,4800)(5700,4800)


\path(2100,1200)(2100,5700)
\path(3000,1200)(3000,5700)
\path(3900,1200)(3900,5700)
\path(4800,1200)(4800,5700)

\put(0500,4800){$x_4$}
\put(0500,3900){$x_3$}
\put(0500,3000){$x_2$}
\put(0500,2100){$x_1$}

\put(2100,0500){$y_1$}
\put(3000,0500){$y_2$}
\put(3900,0500){$y_3$}
\put(4800,0500){$y_4$}

\end{picture}
\label{4-by-4-spin-1/2}
\begin{ca}
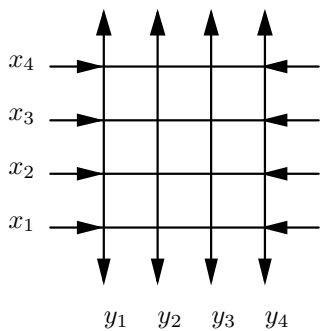
$4\times 4$ spin-$1/2$ partition function.
\end{ca}

\end{minipage}
\end{center}

\end{multicols}

\begin{multicols}{2}

Now, suppose we do not wish to fuse all the way down to the $c+$ 
vertex, which is, the $1\times 1$ DW spin-2 partition function, 
but only {\it half way} to the $2\times 2$ DW spin-1 partition 
function. 


\begin{center}
\begin{minipage}{2in}

\setlength{\unitlength}{0.0008cm}

\begin{picture}(3000, 4500)(-200, 1000)
\thicklines


\blacken\path(1162,2932)(1522,3022)(1162,3112)(1162,2932)
\blacken\path(1162,3832)(1522,3922)(1162,4012)(1162,3832)

\blacken\path(0802,2932)(1162,3022)(0802,3112)(0802,2932)
\blacken\path(0802,3832)(1162,3922)(0802,4012)(0802,3832)


\blacken\path(2762,4012)(2402,3922)(2762,3832)(2762,4012)
\blacken\path(3122,4012)(2762,3922)(3122,3832)(3122,4012)

\blacken\path(2762,3112)(2402,3022)(2762,2932)(2762,3112)
\blacken\path(3122,3112)(2762,3022)(3122,2932)(3122,3112)


\blacken\path(1432,1882)(1522,1522)(1612,1882)(1432,1882)
\blacken\path(1432,2242)(1522,1882)(1612,2242)(1432,2242)

\blacken\path(2332,1882)(2422,1522)(2512,1882)(2332,1882)
\blacken\path(2332,2242)(2422,1882)(2512,2242)(2332,2242)


\blacken\path(1432,5062)(1522,5422)(1612,5062)(1432,5062)
\blacken\path(1432,4702)(1522,5062)(1612,4702)(1432,4702)

\blacken\path(2332,5062)(2422,5422)(2512,5062)(2332,5062)
\blacken\path(2332,4702)(2422,5062)(2512,4702)(2332,4702)

\path(0022,3922)(3922,3922)
\path(0022,3022)(3922,3022)

 \put(-600,3922){$x_2$}
 \put(-600,3022){$x_1$}


\path(1522,1522)(1522,5422) 
\path(2422,1522)(2422,5422)

 \put(1522,1000){$y_1$}
 \put(2422,1000){$y_2$}

\end{picture}

\begin{ca}
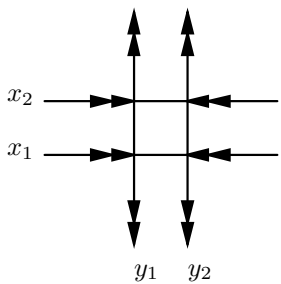
$2\times 2$ Spin-1 partition function.
\end{ca}

\end{minipage}
\end{center}

\end{multicols}

\subsection*{Partial fusion}

It seems plausible that all we need to do in this case is 
to 2-stack the rapidity variables, and normalize suitably. 
In other words, we need {\it partial fusion} as follows. 

\begin{enumerate}

\item Consider the spin-$1/2$ model on a $kL\times kL$ lattice
with DW boundary conditions.

\item Set  
      $\{x_1, $ $x_2, $ $\cdots, $ $x_{kL}\}$ to $L$ $k$-stacks 
      $\{ \{{\bf x_1} | k\},$ 
      $   \{{\bf x_2} | k\},$ $\cdots,$ 
      $   \{{\bf x_L} | k\}\}$
	 and
      $\{y_1, $ $y_2, $ $\cdots, $ $y_{kL}\}$ to $L$ $k$-stacks
      $\{ \{{\bf y_1} | k\},$
      $   \{{\bf y_2} | k\},$ $\cdots,$
      $   \{{\bf y_L} | k\}\}$. 
      Under this restriction of variables, the 
      $M^{1/2\times 1/2}_{kL\times kL}$ IK matrix, that we started 
      with, is now denoted by
      $M^{1/2\times 1/2}_{kL\times kL}$.

\item Normalize so that the weight of the spin-$k/2$ $c+$ vertex 
      is a constant\footnote{The rationale of normalization is to 
      put the result in a practical, recognizable form, and in 
      particular to avoid that the weights have common, spurious 
      poles or zeros.} by dividing with

$$
N^{k/2\times k/2}_{L\times L} 
\ll
\{{\bf x}\}, \{{\bf y}\}
\rr  
= \pl_{1 \le i, j \le L} 
{\pl_{p = 0}^{k-1}[-x_i+ y_j +p]_{k-1}}
$$
\end{enumerate}

Following the above procedure, we obtain the following 
expression for the spin-$k/2$ partition function

\vspace{5mm}

\begin{boxedminipage}[c]{12cm}
\begin{multline}\label{main-result}
Z^{k/2\times k/2}_{L\times L}\ll \{{\bf x}\},\{{\bf y}\}\rr= \\
\frac{
\pl_{1 \le i, j \le L}
\pl_{p = 1}^{k} [-x_i + y_j + p]_{k+1}
}
{\pl_{1 \leq i<j \leq L} 
\pl_{p=0}^{k-1} [-x_i + x_j + p]_{k}
\pl_{p=0}^{k-1} [-y_j + y_i + p]_{k}
} \\
\times det \ll M^{k/2\times k/2}_{kL\times kL}\rr
\end{multline}
\end{boxedminipage}

\vspace{5mm}

Equation \ref{main-result} is our main result. In the next section, 
we show that partial fusion, as outlined above, works, and leads to 
spin-$k/2$ partition functions, with {\it no missing or unwanted} 
configurations. Technical details of how equation \ref{main-result} 
is obtained are in the appendix.

\section{Spin-$k/2$ partition functions: Proof}\label{proof}

We wish to show that, starting from a weighted sum over spin-$1/2$ 
configurations, on a $kL\times kL$ lattice, and dividing the lattice 
into $L^2$ $k\times k$ blocks, we can fuse these blocks one by one, 
and obtain a weighted sum over spin-$k/2$ configurations with the 
correct spin-$k/2$ weights. 

In particular, we also wish to show that This {\it partial} fusion 
procedure leads to {\it all} required spin-$k/2$ configurations, 
{\it and no more}. In fact, it will turn out that this procedure 
is bijective in the sense that every step is reversible.

The following is an outline of the proof, together with a simple 
running example. 

\subsection{Outline of proof}

\begin{enumerate}

\item Consider a DW spin-$1/2$ model on a $kL\times kL$ lattice.
      As an example, we take $k=2$ and $L=3$. 

\item Set the horizontal rapidities into $L$ $k$-stacks of the form 
      $\{
      \{{\bf x_1} | k\},$ $  
      \{{\bf x_2} | k\},$ $  
      \cdots    $ $  
      \{{\bf x_L} | k\}
      \}$, and similarly for the vertical rapidities.
      In our example, we obtain 
      $\{
       \{x_1| 2\}, 
       \{x_2| 2\}, 
       \{x_3| 2\} 
       \}$ 
       and
       $\{
        \{x_1| 2\},
        \{x_2| 2\},
        \{x_3| 2\}
        \}$

\item Divide the lattice into $L^2$ $k\times k$ blocks, 
      $B^{k\times k}_{ij}$, 
      where $i$ is the block row    index, 
            $j$ is the block column index, and 
      $\{i, j\} \in \{1, 2, \cdots, L\}$. Each block has 
      one independent horizontal rapidity $x_i$, and one 
      independent vertical rapidity $y_j$. In our example, 
      we obtain nine $2\times 2$ blocks. 


\item Order the blocks, firstly in terms of row position: 
      blocks with a smaller $i$ precede those with 
      a larger $i$, then in terms of column position: 
      for equal $i$ indices, blocks with a smaller $j$ 
      precede those with larger $j$. 
      In our example, the order is 
      $\{$
      $B^{2\times 2}_{11},$ 
      $B^{2\times 2}_{12},$ 
      $B^{2\times 2}_{13},$ 
      $B^{2\times 2}_{21},$ 
      $B^{2\times 2}_{22},$ 
      $B^{2\times 2}_{23},$
      $B^{2\times 2}_{31}$, 
      $B^{2\times 2}_{32},$ 
      $B^{2\times 2}_{33}$ 
      $\}$

\begin{center}
\begin{minipage}{3in}

\setlength{\unitlength}{0.0008cm}

\begin{picture}(8000, 8000)(0, 1000)
\thicklines

\blacken\path(1815,2622)(2175,2712)(1815,2802)(1815,2622)
\blacken\path(1815,3522)(2175,3612)(1815,3702)(1815,3522)
\blacken\path(1815,4422)(2175,4512)(1815,4602)(1815,4422)
\blacken\path(1815,5322)(2175,5412)(1815,5502)(1815,5322)
\blacken\path(1815,6222)(2175,6312)(1815,6402)(1815,6222)
\blacken\path(1815,7122)(2175,7212)(1815,7302)(1815,7122)

\blacken\path(2085,2172)(2175,1812)(2265,2172)(2085,2172)
\blacken\path(2985,2172)(3075,1812)(3165,2172)(2985,2172)
\blacken\path(3885,2172)(3975,1812)(4065,2172)(3885,2172)
\blacken\path(4785,2172)(4875,1812)(4965,2172)(4785,2172)
\blacken\path(5685,2172)(5775,1812)(5865,2172)(5685,2172)
\blacken\path(6585,2172)(6675,1812)(6765,2172)(6585,2172)

\blacken\path(2265,7677)(2175,8037)(2085,7677)(2265,7677)
\blacken\path(3165,7677)(3075,8037)(2985,7677)(3165,7677)
\blacken\path(4065,7677)(3975,8037)(3885,7677)(4065,7677)
\blacken\path(4965,7677)(4875,8037)(4785,7677)(4965,7677)
\blacken\path(5865,7677)(5775,8037)(5685,7677)(5865,7677)
\blacken\path(6765,7677)(6675,8037)(6585,7677)(6765,7677)

\blacken\path(7035,2802)(6675,2712)(7035,2622)(7035,2802)
\blacken\path(7035,3702)(6675,3612)(7035,3522)(7035,3702)
\blacken\path(7035,4602)(6675,4512)(7035,4422)(7035,4602)
\blacken\path(7035,5502)(6675,5412)(7035,5322)(7035,5502)
\blacken\path(7035,6402)(6675,6312)(7035,6222)(7035,6402)
\blacken\path(7035,7302)(6675,7212)(7035,7122)(7035,7302)

\thinlines 

\dashline{200}(0975,5862)(7875,5862)
\dashline{200}(0975,4062)(7875,4062)


\dashline{200}(3525,1512)(3525,8412)
\dashline{200}(5325,1512)(5325,8412)


\thicklines

\path(1275,2712)(7575,2712)
\path(1275,3612)(7575,3612)
\path(1275,4512)(7575,4512)
\path(1275,5412)(7575,5412)
\path(1275,6312)(7575,6312)
\path(1275,7212)(7575,7212)


\path(2175,2212)(2175,8037)
\path(3075,2212)(3075,8037)
\path(3975,2212)(3975,8037)
\path(4875,2212)(4875,8037)
\path(5775,2212)(5775,8037)
\path(6675,2212)(6675,8037)

\put(0075,7212){$\tilde{x}_3$}
\put(0075,6312){$x_3    $}
\put(0075,5412){$\tilde{x}_2$}
\put(0075,4512){$x_2    $}
\put(0075,3612){$\tilde{x}_1$}
\put(0075,2712){$x_1    $}

\put(2175,1212){$y_1    $}
\put(2775,1212){$\tilde{y}_1$}
\put(3975,1212){$y_2    $}
\put(4575,1212){$\tilde{y}_2$}
\put(5775,1212){$y_3    $}
\put(6375,1212){$\tilde{y}_3$}

\end{picture}

\begin{ca}
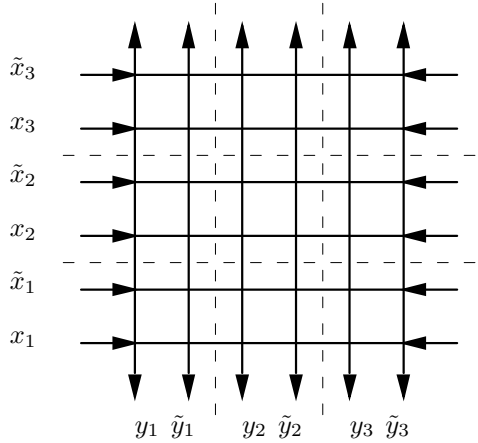
Spin-$1/2$ partition function with 2-stacked rapidity variables.
$\tilde{z} = z+1$.
\end{ca}

\end{minipage}
\end{center}


\item Consider the partition function 
      $Z^{1/2\times 1/2}_{kL\times kL}$ 
      as a sum over products of two partition functions: that of 
      $B^{k\times k}_{11}$, and that of the rest of the lattice, 
      $R_{11}$, that is
      
\begin{equation}\label{first-sum}
Z^{1/2\times 1/2}_{kL\times kL} = \sul_{\{h_{1}, v_{1}\}} 
            B^{k\times k}_{11}
	    R_{11}
\end{equation}
where $h_{1}$ and $v_{1}$ stand for the horizontal and vertical 
common spin boundaries. The sum is over all configurations on 
the common boundaries of $B^{k\times k}_{11}$ and $R_{11}$. 

\item Each $B^{k\times k}_{11}$ term in the above sum is DW on 
      an inflow boundary, and has a single $\sigma$-configuration 
      on an outflow boundary. Since 
      a DW boundary condition corresponds to a $\sigma$-set 
      with a single $\sigma$-configuration, the inflow boundaries 
      are (trivially) 
      summed, while the outflow boundaries are fixed to a certain 
      $\sigma$-configuration. But, from fusion, all such partition 
      functions are equal to the one with 
      $\sigma$-representatives on the outflow boundaries. This 
      allows us to simplify the sum in the previous equation to
\begin{equation}\label{secondsum}
Z^{1/2\times 1/2}_{L\times L} = \sum
       \ll B^{k\times k}_{11} \sum R_{11} \rr
\end{equation}
\noindent where the first (right most) sum is over all 
$\sigma$-configurations, in an allowed $\sigma$-set on the inflow 
boundaries of $R_{11}$, $B^{k\times k}_{11}$ has 
$\sigma$-representatives on the outflow boundaries, and the second 
(left most) sum is over all $\sigma$-sets on the common boundaries.
Notice that we need to be clear about what is being summed over in 
each sum. This is because, before fusion, all elements in all
allowed $\sigma$-sets are summed over, while, after fusion, only
$\sigma$-representatives are summed over.

%
\begin{center}
\begin{minipage}{5in}
\setlength{\unitlength}{0.0008cm}
\begin{picture}(5000, 8000)(0, 0)
\thicklines
%
\blacken\path(0315,0564)(0675,0654)(0315,0744)(0315,0564)
\blacken\path(0315,1464)(0675,1554)(0315,1644)(0315,1464)
%
\blacken\path(0585,0414)(0675,0054)(0765,0414)(0585,0414)
\blacken\path(1485,0414)(1575,0054)(1665,0414)(1485,0414)
%
\blacken\path(1815,3264)(2175,3354)(1815,3444)(1815,3264)
\blacken\path(1815,4164)(2175,4254)(1815,4344)(1815,4164)
\blacken\path(1815,5064)(2175,5154)(1815,5244)(1815,5064)
\blacken\path(1815,5964)(2175,6054)(1815,6144)(1815,5964)
%
\blacken\path(2265,6594)(2175,6954)(2085,6594)(2265,6594)
\blacken\path(3165,6594)(3075,6954)(2985,6594)(3165,6594)
%
%
%
\blacken\path(3885,1014)(3975,0654)(4065,1014)(3885,1014)
\blacken\path(4065,6594)(3975,6954)(3885,6594)(4065,6594)
\blacken\path(4785,1014)(4875,0654)(4965,1014)(4785,1014)
\blacken\path(4965,6594)(4875,6954)(4785,6594)(4965,6594)
\blacken\path(5685,1014)(5775,0654)(5865,1014)(5685,1014)
\blacken\path(5865,6594)(5775,6954)(5685,6594)(5865,6594)
\blacken\path(6585,1014)(6675,0654)(6765,1014)(6585,1014)
\blacken\path(6765,6594)(6675,6954)(6585,6594)(6765,6594)
%
%
\blacken\path(7035,1644)(6675,1554)(7035,1464)(7035,1644)
\blacken\path(7035,3444)(6675,3354)(7035,3264)(7035,3444)
\blacken\path(7035,5244)(6675,5154)(7035,5064)(7035,5244)
\blacken\path(7035,6144)(6675,6054)(7035,5964)(7035,6144)
\blacken\path(7035,2544)(6675,2454)(7035,2364)(7035,2544)
\blacken\path(7035,4344)(6675,4254)(7035,4164)(7035,4344)
%
%
%
\path(0075,0654)(2175,0654)
\path(0075,1554)(2175,1554)
\put(-1075,1554){$\tilde{x}_1$}
\put( 2375,1554){$v_1    $}

\put(-1075,0654){$x_1    $}
\put( 2375,0654){$v_1    $}
%
%
%
\path(0675,0054)(0675,2154)
\path(1575,0054)(1575,2154)
\put(0875,0000){$y_1    $}
\put(0875,2000){$h_1    $}

\put(1775,0000){$\tilde{y}_1$}
\put(1775,2000){$h_1    $}
%
%
%
\dashline{200}(0975,4704)(7875,4704)
\dashline{200}(0975,2904)(7875,2904)
%
%
%
\path(1275,3354)(7575,3354)
\path(1275,4254)(7575,4254)
\path(1275,5154)(7575,5154)
\path(1275,6054)(7575,6054)
%
%
%
\path(2175,2754)(2175,6954)
\path(3075,2754)(3075,6954)
%
%
\path(3375,1554)(7575,1554)
\path(3375,2454)(7575,2454)
%
%
%
\dashline{200}(3525,0354)(3525,7254)
\dashline{200}(5325,0354)(5325,7254)
%
%
%
\path(3975,0654)(3975,6954)
\path(4875,0654)(4875,6954)
\path(5775,0654)(5775,6954)
\path(6675,0654)(6675,6954)
\put(0075,6054){$\tilde{x}_3$}
\put(0075,5154){$x_3    $}
\put(0075,4254){$\tilde{x}_2$}
\put(0075,3354){$x_2    $}
\put(3775,0054){$y_2    $}
\put(4575,0054){$\tilde{y}_2$}

\put(5675,0054){$y_3    $}
\put(6475,0054){$\tilde{y}_3$}
\end{picture}
\begin{ca}
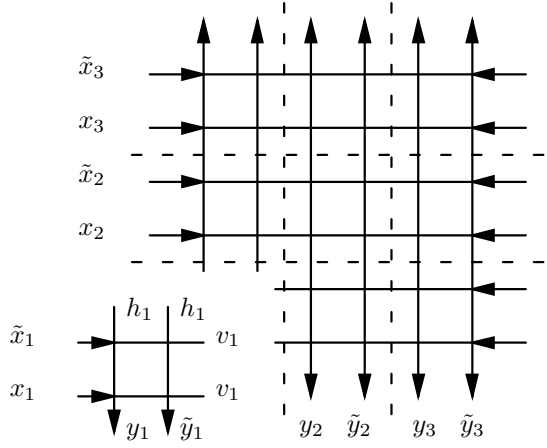
Detaching a $2\times 2$ spin-$1/2$ block. $\tilde{z} = z+1$.
\end{ca}
\end{minipage}
\end{center}

\item Using fusion to write $B^{k\times k}_{11}$ as a spin-$k/2$ 
vertex, we end up with a sum over products of a $1\times 1$ 
spin-$k/2$ non-DW partition function and a spin-$1/2$ non-DW 
partition function (the original lattice minus $B^{k\times k}_{11}$).

\begin{equation}\label{thirdsum}
Z^{1/2\times 1/2}_{kL\times kL} = \sum
      \ll
      Z^{k/2\times k/2}_{1 \times 1} 
      \sum 
      Z^{1/2\times 1/2}_{(kL \times kL) - (1 \times 1)}
      \rr
\end{equation}

%
\begin{center}
\begin{minipage}{4in}
\setlength{\unitlength}{0.0008cm}
\begin{picture}(5000, 8000)(-200, 0)
\thicklines
%
%
\blacken\path(0765,1057)(1125,1147)(0765,1237)(0765,1057)
\blacken\path(0405,1057)(0765,1147)(0405,1237)(0405,1057)

\put(-0475,1057){$x_1$}
%
%
\blacken\path(1035,0382)(1125,0022)(1215,0382)(1035,0382)
\blacken\path(1035,0742)(1125,0382)(1215,0742)(1035,0742)

\put(1435,0082){$y_1$}
%
%
%
\blacken\path(1815,3757)(2175,3847)(1815,3937)(1815,3757)
\blacken\path(1815,4657)(2175,4747)(1815,4837)(1815,4657)
\blacken\path(1815,5557)(2175,5647)(1815,5737)(1815,5557)
\blacken\path(1815,6457)(2175,6547)(1815,6637)(1815,6457)
%
%
\blacken\path(2265,7087)(2175,7447)(2085,7087)(2265,7087)
\blacken\path(3165,7087)(3075,7447)(2985,7087)(3165,7087)
%
%
\blacken\path(3885,1507)(3975,1147)(4065,1507)(3885,1507)
\blacken\path(4065,7087)(3975,7447)(3885,7087)(4065,7087)
\blacken\path(4785,1507)(4875,1147)(4965,1507)(4785,1507)
\blacken\path(4965,7087)(4875,7447)(4785,7087)(4965,7087)
\blacken\path(5685,1507)(5775,1147)(5865,1507)(5685,1507)
\blacken\path(5865,7087)(5775,7447)(5685,7087)(5865,7087)
\blacken\path(6585,1507)(6675,1147)(6765,1507)(6585,1507)
\blacken\path(6780,7087)(6675,7447)(6570,7087)(6780,7087)
%
%
\blacken\path(7035,2137)(6675,2047)(7035,1957)(7035,2137)
\blacken\path(7035,3037)(6675,2947)(7035,2857)(7035,3037)
\blacken\path(7035,3937)(6675,3847)(7035,3757)(7035,3937)
\blacken\path(7035,4837)(6675,4747)(7035,4657)(7035,4837)
\blacken\path(7035,5737)(6675,5647)(7035,5557)(7035,5737)
\blacken\path(7035,6637)(6675,6547)(7035,6457)(7035,6637)
%
%
%
\path(0075,1147)(2175,1147)
%
%
\path(1125,0022)(1125,2122)
%
\dashline{200}(0975,5197)(7875,5197)
\dashline{200}(0975,3397)(7875,3397)
\dashline{200}(3525,0847)(3525,7747)
\dashline{200}(5325,0847)(5325,7747)
%
%
%
\path(1275,3847)(7575,3847)
\path(1275,4747)(7575,4747)
\path(1275,5647)(7575,5647)
\path(1275,6547)(7575,6547)
%
%
\path(2175,3247)(2175,7447)
\path(3075,3247)(3075,7447)
%
%
\path(3375,2047)(7575,2047)
\path(3375,2947)(7575,2947)
%
%
\path(3975,1147)(3975,7447)
\path(4875,1147)(4875,7447)
\path(5775,1147)(5775,7447)
\path(6675,1147)(6675,7447)
%
%
\put(0075,6547){$\tilde{x}_3$}
\put(0075,5647){$x_3    $}
\put(0075,4747){$\tilde{x}_2$}
\put(0075,3847){$x_2    $}
\put(3975,0547){$y_2    $}
\put(4675,0547){$\tilde{y}_2$}
\put(5775,0547){$y_3    $}
\put(6675,0547){$\tilde{y}_3$}
\end{picture}
\begin{ca}
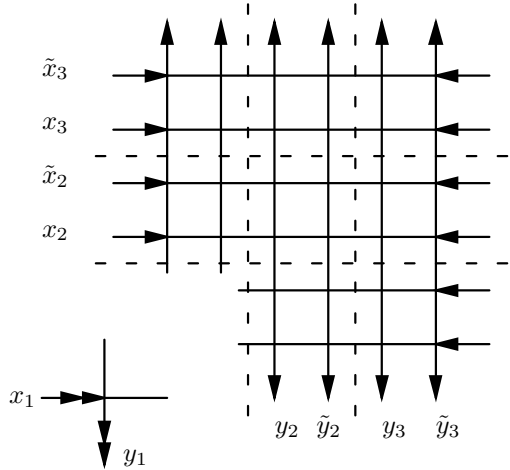
Fusing a detached $2\times 2$ spin-$1/2$ block into a $1\times 1$ spin-1
partition function. $\tilde{z} = z+1$.
\end{ca}
\end{minipage}
\end{center}

\item Next, we consider the next ranking block in 
      $Z^{1/2\times 1/2}_{(kL \times kL) - (1 \times 1)}$, 
      namely $B^{k\times k}_{12}$, and write

\begin{equation}\label{fourthsum}
Z^{1/2\times 1/2}_{(kL \times kL) - (1 \times 1)} = 
\sum
B^{k\times k}_{12}
Z^{1/2\times 1/2}_{(kL \times kL) - (2 \times 1)}	    
\end{equation}

Using the same reasoning, and notation, as above, we can write

\begin{equation}\label{fifthsum}
Z^{1/2\times 1/2}_{(kL \times kL) - (1 \times 1)} =
\sum
\ll
B^{k\times k}_{12}
\sum
Z^{1/2\times 1/2}_{(kL \times kL) - (2 \times 1)}
\rr
\end{equation}

\begin{multicols}{2}

Using fusion to re-write $B^{k\times k}_{12}$ as a spin-$k/2$ 
vertex, combining the above results, and summing over the common 
boundaries of the two $1 \times 1$ spin-$k/2$ partition functions, 
we can write the initial spin-$1/2$ non-DW partition function as 
a sum over products of two objects: a spin-$k/2$ non-DW partition 
function consisting of 2 vertices, and the remaining spin-$1/2$ 
lattice. 

It should be clear from the above that we are fusing the initial 
spin-$1/2$ lattice, one block at a time, to a spin-$k/2$ lattice.
It should also be clear that this can be done block by block, that 
the procedure is reversible, and that we end up with the desired 
$L\times L$ DW spin-$k/2$ model. 

The DW boundary conditions of the final, spin-$k/2$ configurations 
follow from the DW boundary conditions of the initial spin-$1/2$ 
configurations.


\begin{center}
\begin{minipage}{2.5in}

\setlength{\unitlength}{0.0008cm}

\begin{picture}(3000, 8000)(-500, 0)
\thicklines


\blacken\path(0555,1132)(0915,1222)(0555,1312)(0555,1132)
\blacken\path(0915,1132)(1275,1222)(0915,1312)(0915,1132)


\blacken\path(1185,0382)(1275,0022)(1365,0382)(1185,0382)
\blacken\path(1185,0742)(1275,0382)(1365,0742)(1185,0742)


\blacken\path(2085,0742)(2175,0382)(2265,0742)(2085,0742)
\blacken\path(2085,0382)(2175,0022)(2265,0382)(2085,0382)


\blacken\path(1815,4132)(2175,4222)(1815,4312)(1815,4132)
\blacken\path(1815,5032)(2175,5122)(1815,5212)(1815,5032)
\blacken\path(1815,5932)(2175,6022)(1815,6112)(1815,5932)
\blacken\path(1815,6832)(2175,6922)(1815,7012)(1815,6832)


\blacken\path(2265,7462)(2175,7822)(2085,7462)(2265,7462)
\blacken\path(3165,7462)(3075,7822)(2985,7462)(3165,7462)
\blacken\path(4065,7462)(3975,7822)(3885,7462)(4065,7462)
\blacken\path(4965,7462)(4875,7822)(4785,7462)(4965,7462)
\blacken\path(5865,7462)(5775,7822)(5685,7462)(5865,7462)
\blacken\path(6765,7462)(6675,7822)(6585,7462)(6765,7462)


\blacken\path(5685,1882)(5775,1522)(5865,1882)(5685,1882)
\blacken\path(6585,1882)(6675,1522)(6765,1882)(6585,1882)


\blacken\path(7035,2512)(6675,2422)(7035,2332)(7035,2512)
\blacken\path(7035,3412)(6675,3322)(7035,3232)(7035,3412)
\blacken\path(7035,4312)(6675,4222)(7035,4132)(7035,4312)
\blacken\path(7035,5212)(6675,5122)(7035,5032)(7035,5212)
\blacken\path(7035,6112)(6675,6022)(7035,5932)(7035,6112)
\blacken\path(7035,7012)(6675,6922)(7035,6832)(7035,7012)



\path(0075,1222)(3075,1222)
\put(-0475,1222){$x_1$}


\path(1275,0022)(1275,2122)
\path(2175,0022)(2175,2122)

\put(1375,-200){$y_1$}
\put(2275,-200){$y_2$}


\dashline{200}(0975,3772)(7875,3772)
\dashline{200}(0975,5572)(7875,5572)


\dashline{200}(3525,3022)(3525,8122)
\dashline{200}(5325,3022)(5325,8122)


\path(1275,4222)(7575,4222)
\path(1275,5122)(7575,5122)
\path(1275,6022)(7575,6022)
\path(1275,6922)(7575,6922)


\path(2175,3622)(2175,7822)
\path(3075,3622)(3075,7822)
\path(3975,3622)(3975,7822)
\path(4875,3622)(4875,7822)


\path(5775,1522)(5775,7822)
\path(6675,1522)(6675,7822)


\path(5175,3322)(7575,3322)
\path(5175,2422)(7575,2422)


\put(0075,6847){$\tilde{x}_3$}
\put(0075,6022){$x_3    $}
\put(0075,5122){$\tilde{x}_2$}
\put(0075,4222){$x_2    $}

\put(5775,0922){$y_3    $}
\put(6675,0922){$\tilde{y}_3$}

\end{picture}

\begin{ca}
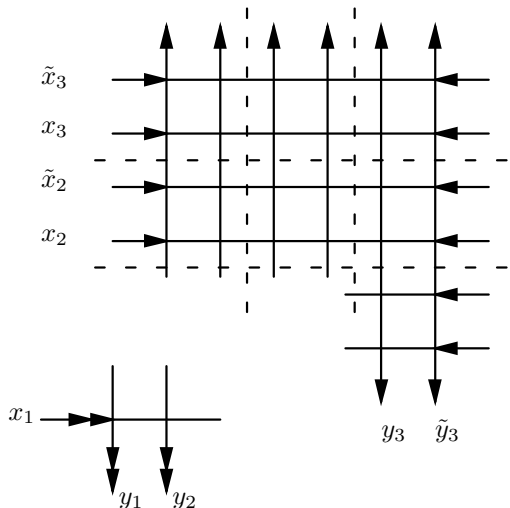
Detaching, fusing the second $2\times 2$ spin-$1/2$ block, fusing
it to form a spin-1 vertex, then attaching the latter to the
first spin-1 vertex. $\tilde{z} = z+1$.
\end{ca}

\end{minipage}
\end{center}

\end{multicols}

\end{enumerate}

This concludes our outline of the proof of equation \ref{main-result}, 
from which the reader can recover a formal proof if necessary. 

\section{Example: Spin-1 model}\label{example}

Consider the spin-1 (19-vertex, or Zamolodchikov-Fateev model
\cite{fz}) 
model constructed by fusion \footnote{There are many 19-vertex 
models. The Zamolodchikov-Fateev model that we are considering
here is only on of these. Two others are listed in 
\cite{lima-santos}.}. The vertex weights, which can be computed 
explicitly using fusion, are listed (for example) in 
\cite{lima-santos}. In the following, we list only those 
vertices that we need and their the weights. Vertices that are 
related to those listed, by inverting of all arrows, have the 
same weights.


\begin{center}
\begin{minipage}{4in}

\setlength{\unitlength}{0.0008cm}
\begin{picture}(10000,6000)(0, 1000)
\thicklines




\blacken\path(1102,5610)(1462,5700)(1102,5790)(1102,5610)
\blacken\path(1462,5610)(1822,5700)(1462,5790)(1462,5610)

\blacken\path(5002,5610)(5362,5700)(5002,5790)(5002,5610)
\blacken\path(5362,5610)(5722,5700)(5362,5790)(5362,5610)

\blacken\path(8902,5610)(9262,5700)(8902,5790)(8902,5610)
\blacken\path(9262,5610)(9622,5700)(9262,5790)(9262,5610)


\blacken\path(1912,6240)(1822,6600)(1732,6240)(1912,6240)
\blacken\path(1912,5880)(1822,6240)(1732,5880)(1912,5880)

\blacken\path(5812,6240)(5722,6600)(5622,6240)(5812,6240)
\blacken\path(5812,5880)(5722,6240)(5622,5880)(5812,5880)

\blacken\path(9712,6240)(9622,6600)(9522,6240)(9712,6240)
\blacken\path(9712,5880)(9622,6240)(9522,5880)(9712,5880)




\blacken\path(1912,5340)(1822,5700)(1732,5340)(1912,5340)
\blacken\path(1912,4980)(1822,5340)(1732,4980)(1912,4980)


\blacken\path(9712,5520)(9622,5160)(9532,5520)(9712,5520)
\blacken\path(9712,5160)(9622,4800)(9532,5160)(9712,5160)


\blacken\path(2362,5610)(2722,5700)(2362,5790)(2362,5610)
\blacken\path(2002,5610)(2362,5700)(2002,5790)(2002,5610)


\blacken\path(09982,5610)(9622,5700)(09982,5790)(09982,5610)
\blacken\path(10342,5610)(9982,5700)(10342,5790)(10342,5610)



\blacken\path(1912,2940)(1822,3300)(1732,2940)(1912,2940)
\blacken\path(1912,2580)(1822,2940)(1732,2580)(1912,2580)

\blacken\path(1912,2040)(1822,2400)(1732,2040)(1912,2040)
\blacken\path(1912,1680)(1822,2040)(1732,1680)(1912,1680)

\blacken\path(1282,2310)(0922,2400)(1282,2490)(1282,2310)
\blacken\path(1642,2310)(1282,2400)(1642,2490)(1642,2310)

\blacken\path(2182,2310)(1822,2400)(2182,2490)(2182,2310)
\blacken\path(2542,2310)(2182,2400)(2542,2490)(2542,2310)


\blacken\path(5812,2940)(5722,3300)(5632,2940)(5812,2940)
\blacken\path(5812,2580)(5722,2940)(5632,2580)(5812,2580)

\blacken\path(6082,2310)(5722,2400)(6082,2490)(6082,2310)
\blacken\path(6442,2310)(6082,2400)(6442,2490)(6442,2310)


\blacken\path(9712,2940)(9622,3300)(9532,2940)(9712,2940)
\blacken\path(9712,2580)(9622,2940)(9532,2580)(9712,2580)

\blacken\path(9712,2040)(9622,2400)(9532,2040)(9712,2040)
\blacken\path(9712,1680)(9622,2040)(9532,1680)(9712,1680)






\path(0922,2400)(2722,2400)  
\path(0922,5700)(2722,5700)  

\path(4822,2400)(6622,2400)  
\path(4822,5700)(6622,5700)  

\path(8722,2400)(10522,2400)  
\path(8722,5700)(10522,5700)  



\path(1822,1500)(1822,3300)  
\path(1822,4800)(1822,6600)  

\path(5722,1500)(5722,3300)  
\path(5722,4800)(5722,6600)  

\path(9622,1500)(9622,3300)  
\path(9622,4800)(9622,6600)  








\put(0800,4200){$A=[u+1][u+2]$}
\put(4800,4200){$X=[u+1][2]$}
\put(8800,4200){$C=[1][2]$}

\put(0800,0900){$B=[u-1][u]$}
\put(4800,0900){$Y=[u][2]$}
\put(8800,0900){$E=[u][u+1]$}

\end{picture}

\begin{ca}
\label{nineteenvertices}
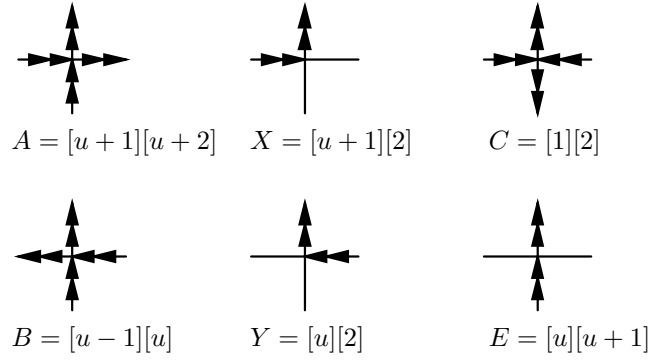
A subset of the vertices of the spin-$1$ model and their weights. 
$u=-x+y$.
\end{ca}

\end{minipage}
\end{center}


Re-writing equation \ref{main-result}, for $k=2$, in terms of vertex 
weights of the spin-1 model, we obtain the following expression for 
the spin-1 partition function

\begin{multline}
\label{spin-1-dw}
Z^{1\times 1}_{L\times L} 
\ll 
\{{\bf x}\},\{{\bf y}\}
\rr=  \\
\frac{
\pl_{1 \le i, j \le L} 
A(-x_i+ y_j) E(-x_i+ y_j) B(-x_i+ y_j)
}
{\pl_{i < j} 
E(-x_i+ x_j) B(-x_i+ x_j) 
E( y_i- y_j) B( y_i- y_j)
}\\
\times det \ll M^{2\times 2}_{2L\times 2L}\rr 
\end{multline}

\begin{align}
\begin{array}{lll}
M_{2j-1,             2i-1}^{2\times 2}             ={1}/{E(-x_i+y_j)},&
M_{2j-1,             2i{\phantom{-1}}}^{2\times 2} ={1}/{B(-x_i+y_j)},\\
\phantom{space}&\phantom{space}\\
M_{2j,{\phantom{-1}} 2i-1}^{2\times 2}             ={1}/{A(-x_i+y_j)},&
M_{2j,{\phantom{-1}} 2i{\phantom{-1}}}^{2\times 2} ={1}/{E(-x_i+y_j)}
\end{array}
\end{align}

\subsection*{Independent check} The determinant expression of the 
spin-1 partition function, obtained above, allows for an independent 
check of our general result, in the sense that one can take the 
determinant expression as a conjecture, and show that it is correct. 

Just as in the spin-$1/2$ case \cite{korepin,izergin}, one needs 
to show that the LHS of equation \ref{spin-1-dw} (the partition function) 
has certain properties that uniquely determine it completely, then show 
that the RHS (the proposed determinant expression) satisfies the same 
properties. An outline of the basic steps is as follows. 

\subsection*{The LHS of equation \ref{spin-1-dw}} The following 
properties fully characterize the partition function 

\noindent {\bf Symmetry} Using the Yang Baxter equations, one can
show that the partition function is a symmetric function in 
$\{{\bf x}\}$ and in $\{{\bf y}\}$. 

\noindent {\bf Degree} One can easily show that, on any extremal  
row or column, there is exactly one rapidity independent vertex 
(namely a $c+$ vertex), {\it or} two vertices that are trigonometric 
polynomials of degree 1 each, while all other vertices are of degree 2. 
This means that the partition function is a trigonometric polynomial 
of degree $2L-2$ in the rapidity variable in that row or column, and 
by symmetry in every other variable. 

\noindent {\bf Recursion relations} Below, we will show that 
the partition function satisfies $2L$ recursion relations in 
each rapidity variable, which is more than we actually need.

\noindent {\bf The initial condition} By construction, the $1\times 1$ 
DW partition is a $c+$ vertex.

\subsection*{The RHS of equation \ref{spin-1-dw}} Next we show
that the RHS satisfies the same properties as the LHS.

\noindent {\bf Symmetry} By direct calculation, one can show that the 
RHS is a symmetric function in $\{{\bf x}\}$ and in $\{{\bf y}\}$. 

\noindent {\bf Degree} Naive power counting shows that the RHS is 
a degree $2L$ trigonometric polynomial in any rapidity variable
$x$. However, taking the limit $x \rightarrow \infty$, for real 
crossing parameter, one can show explicitly that degree $2L$ terms 
cancel, while there are no degree $2L-1$ terms, so that the RHS is  
a degree $2L-2$ trigonometric polynomial.

\noindent {\bf Recursion relations} The RHS satisfies the same 
recursion relations as LHS, as will be shown below.

\noindent {\bf Initial condition} By direct calculation, the RHS 
for $L=1$ reduces to the weight of the $c+$ vertex. 

\subsection*{Recursions from upper left corner} 
Due to the boundary conditions, the only vertices that are allowed 
at the upper left corner are
$A(-x_1+y_1)$, 
$X(-x_1+y_1)$ and  
$C(-x_1+y_1)$. 

Setting $-x_1+y_1+1=0$, we obtain $A(-1)=X(-1)=0$, which 
freezes all vertices on the top row and first column and 
leads to the recursion relation

\begin{multline}
Z^{1\times 1}_{L\times L}
\ll 
\{{\bf x}\},\{{\bf y}\} | x_1=y_1+1
\rr = \\
[1][2]
\ll \pl_{j=2}^L B(-x_1+y_{j-1}) B(-x_j+y_1)\rr 
Z^{1\times 1}_{(L-1)\times (L-1)}
\ll \{ {\bf x} \}_{1}, \{ {\bf y} \}_{1} \rr
\end{multline}  

Given the symmetry in vertical rapidities, we get the same 
relations for $x_i=y_j+1$, for all $i$ and all $j$, so we have 
$L$ recursion relations for each rapidity variable.

\subsection*{Recursions from upper right corner} 
Due to the boundary conditions, the only vertices that 
are allowed at the upper right corner are 
$B(-x_1+y_L)$, 
$Y(-x_1+y_L)$ and 
$C(-x_1+y_L)$. 

Setting $-x_1+y_1=0$, we obtain $B(0)=Y(0)=0$, and only $C(0)$ 
survives at the corner, freezing all the vertices on the top row 
and last column. The remaining $(L-1)\times (L-1)$ lattice has 
once again DW boundary conditions, and we obtain the recursion 
relation

\begin{multline}
Z^{1\times 1}_{L\times L}
\ll 
\{{\bf x}\},\{{\bf y}\} | x_1=y_L 
\rr =\\
[1][2] 
\ll \pl_{j=2}^L A(-x_1+y_j) A(-x_j+y_L)\rr
Z^{1\times 1}_{(L-1)\times (L-1)}\ll\{{\bf x}\}_{1}, \{{\bf y}\}_{L}\rr
\end{multline}   

Given the symmetry in vertical rapidities, a similar relation, 
$x_i=y_j$, can be written for all $i$ and $j$, and we have $L$ 
recursion relations for each rapidity variable. Thus we have 
altogether $2L$ recursion relations for each variable, which 
are sufficient to completely determine the partition function 
as a trigonometric polynomial of degree $2L-2$ in that variable, 
just as Izergin's proof.

As the RHS of equation \ref{spin-1-dw} satisfies the $2L$ required 
recursion relations and is a polynomial of degree $2L-2$, in every 
rapidity variable, and given that the initial condition is satisfied, 
we conclude that it coincides with the LHS of equation \ref{spin-1-dw}.

\subsection*{Comments} The above Izergin type proof does not extend 
to higher spin models, beyond spin-1. The reason is that, for $k > 2$, 
the degree of the polynomials that we need to determine is higher 
than the number of available recursion relations\footnote{This assumes 
that we are willing to consider only Lagrange interpolation in determining 
the polynomials under consideration. One can consider more elaborate 
interpolations, such as Hermite interpolation, but then things become 
very complicated.}.

\section{The homogeneous Limit}\label{hom-lim}

Taking the homogeneous limit of the spin-$k/2$ partition function, 
following the footsteps of \cite{izergin}, is straightforward. For 
convenience, we re-write $M^{k/2\times k/2}_{kL\times kL, (i,j)}$ as

\begin{multline}
M^{k/2\times k/2}_{kL\times kL, (i,j)}=\\
\ll
\begin{array}{llll}
\phi(-x_{i}+y_{j})&\phi(-x_{i}+y_{j}+1)&\ldots&\phi(-x_{i}+y_{j}+k-1) \\ 
\phi(-x_{i}+y_{j}-1)&\phi(-x_{i}+y_{j})&&\vdots \\ 
\vdots&&\ddots& \\ 
\phi(-x_{i}+y_{j}-k+1)&\ldots&&\phi(-x_{i}+y_{j}) 
\end{array}
\rr 
\nonumber
\end{multline}

\noindent where $\phi(x)=1/([x][x+1])$. Let $x_{1}=x$ and consider 
the limit $x_{2}\rightarrow x$. The first block row 
$M^{k/2\times k/2}_{kL\times kL, 1j}$ remains unchanged apart from 
replacing $x_{1}$ with $x$, while $M^{k/2\times k/2}_{kL\times kL, 2j}$ 
becomes 

\begin{multline}
M^{k/2\times k/2}_{kL\times kL, (2,j)}= \\
\ll
\begin{array}{llll}
\phi(-x_{2}+y_{j})&\phi(-x_{2}+y_{j}+1)&\ldots&\phi(-x_{2}+y_{j}+k-1) \\ 
\phi(-x_{2}+y_{j}-1)&\phi(-x_{2}+y_{j})&&\vdots \\ 
\vdots&&\ddots& \\ 
\phi(-x_{2}+y_{j}-k+1)&\ldots&&\phi(-x_{2}+y_{j}) 
\end{array}
\rr 
\nonumber
\end{multline}

Taylor expanding each term as $x_{2}\rightarrow x$, to first
order, and subtracting the first block row from the second 

\begin{align}
(x_{2}-x)^{k}
\ll
\begin{array}{llll}
\phi'(-x+y_{j})&\phi'(-x+y_{j}+1)&\ldots&\phi'(-x+y_{j}+k-1) \\ 
\phi'(-x+y_{j}-1)&\phi'(-x+y_{j})&&\vdots \\ \vdots&&\ddots& \\ 
\phi'(-x+y_{j}-k+1)&\ldots&&\phi'(-x+y_{j}) 
\end{array}
\rr
\nonumber
\end{align}

\noindent where $\phi^{(n)}(x)$ represents the $n^{\rm th}$ derivative 
of $\phi$ with respect to its argument. Note that the $[-x_{2}+x]^{k}$ 
term in the denominator of equation \ref{main-result} cancels exactly 
with the overall factor $(x_{2}-x)^{k}$ as $x_{2}\rightarrow x$. 

Taylor expanding as $x_{i}\rightarrow x$, up to $n$-th order, and 
successively eliminating terms by subtracting multiples of previous 
block rows and taking out common factors, 
$M^{k/2\times k/2}_{kL\times kL, ij}$ becomes 

\begin{multline}
(x_{i}-x)^{k(i-1)} {\ll {(i-1)!} \rr}^{-k}
\times \\
\ll
\begin{array}{llll}
\phi^{(i-1)}(-x+y_{j})&\phi^{(i-1)}(-x+y_{j}+1)&\ldots&
\phi^{(i-1)}(-x+y_{j}+k-1) \\ 
\phi^{(i-1)}(-x+y_{j}-1)&\phi^{(i-1)}(-x+y_{j})&&\vdots \\ 
\vdots&&\ddots& \\ 
\phi^{(i-1)}(-x+y_{j}-k+1)&\ldots&&\phi^{(i-1)}(-x+y_{j}) 
\end{array}
\rr 
\nonumber
\end{multline}

As before, the denominator in equation \ref{main-result} contributes 
a factor of $[-x_{i}+x]^{k(i-1)}$ which cancels with the above 
coefficient as $x_{i}\rightarrow x$, leaving a factor of 
${\ll (i-1)!\rr}^{-k}$, and semi-homogeneous spin-$k/2$ 
$L\times L$ partition function becomes

\begin{multline}
\frac{
{\ll 
\pl_{j=1}^{L}
\pl_{p=0}^{k} 
[-x+y_{j}+p]_k
\rr}^{L}
det \ll M^{k/2\times k/2}_{kL\times kL}\rr
}
{\pl_{i=1}^{k-1}
{\ll -[i]^{2}\rr}^{(k-i)(L^{2}-L)/2}
\pl_{i=1}^{L-1}
{\ll i! \rr}^{k}
\pl_{1\leq i<j \leq L}
\pl_{p=1}^{k} [-y_{i}+y_{j}+p]_k} \\
\label{xhom}
\end{multline}

With $M^{k/2\times k/2}_{kL\times kL, ij}$ given by the above matrix. 
Equation \ref{xhom} is the partition function for a lattice where the 
horizontal rapidities are homogeneous but the vertical rapidities are 
still distinct. Applying similar arguments to the vertical rapidities, 
and combining results, we obtain the following expression for the 
homogeneous $L\times L$ spin-$k/2$ partition function

\begin{equation}
{\mathcal Z}^{k/2\times k/2}_{L\times L}= 
\frac{
{\ll
\pl_{p=0}^{k} [-x+y+p]_k
\rr}^{L^{2}}
det \ll M^{k/2\times k/2}_{kL\times kL}\rr
}
{\pl_{i=1}^{k-1}
\ll
-[i]^{2}
\rr^{(k-i)(L^{2}-L)}
\pl_{i=1}^{L-1}
{\ll i!  \rr}^{2k}
} 
\label{khom}
\end{equation}

\noindent where  
$M^{k/2\times k/2}_{L\times L, ij}$ is 

\begin{align}
\ll
\begin{array}{llll}
\phi^{(i+j-2)}(u)&\phi^{(i+j-2)}(u+1)&\ldots&
\phi^{(i+j-2)}(u+k-1) \\ 
\phi^{(i+j-2)}(u-1)&\phi^{(i+j-2)}(u)&&\vdots \\ 
\vdots&&\ddots& \\ 
\phi^{(i+j-2)}(u-k+1)&\ldots&&\phi^{(i+j-2)}(u) 
\end{array}
\rr 
\nonumber
\end{align}

\subsection*{Remarks on combinatorics} Do the higher spin
determinants lead to interesting combinatorics, as in the 
spin-$1/2$ case? It is not difficult to show that there is 
a simple bijection between spin-$k/2$ DW configurations and 
extended alternating sign matrices ASM's with entries in 
$\{0,$ $ \pm 1,$$ \cdots,$$ \pm k\}$, and conditions that 
naturally extend those of the usual ASM's \cite{fkow}. 

However, one can also easily check that, unlike the spin-1/2 
case, for $k\geq 2$ there is no choice of crossing parameter 
and rapidity variables such that all weights are equal (even 
up to phases). This rules out 1-enumerations (but not weighted 
enumerations). This conclusion is substantiated by direct 
numerical enumerations (for $k=2$), that lead to numbers 
that cannot be expressed as simple products \cite{fkow}.  
We hope to return to these issues in a separate publication.

\section*{Appendix: Technical details}\label{appendix-1}

We start from the partition function of the spin-$1/2$ model
on a $kL\times kL$ lattice

\begin{equation}
Z^{1/2\times 1/2}_{kL\times kL}=
\frac{\pl_{1\leq i,j\leq kL}[-x_{i}+y_{j}+1][-x_{i}+y_{j}]}
     {\pl_{1\leq i<j\leq kL}[-x_{i}+x_{j}] 
      \pl_{1\leq j<i\leq kL}[-y_{i}+y_{j}]}
     det \ll M^{1/2\times 1/2}_{kL\times kL} \rr 
\label{starting-point}
\end{equation}

To obtain the fused partition function, we proceed in two steps. 
Firstly, we set the rapidities to suitable values, then we normalize 
the result. Setting the rapidities to 
$\{ \{{\bf x_1} | k\},$ $ \{{\bf x_2} | k\}, $ $\cdots, $
$   \{{\bf x_L} | k\}
 \}$, 
and
$\{ 
 \{{\bf y_1} | k\},$ $ 
 \{{\bf y_2} | k\},$ $\cdots, $
$\{{\bf y_L} | k\}
\}$,
and using 
$u_{ij} = -x_{i}+y_{j}$, and 
$v_{ij} = -x_{i}+x_{j}$, we obtain

\begin{multline}
\pl_{1\leq i, j \leq kL} [u_{ij}] \rightarrow \\
\pl_{1 \leq i, j \leq L}
[u_{ij}-k+1][u_{ij}-k+2]^{2}
\cdots 
[u_{ij}]^{k}
\cdots 
[u_{ij}+k-2]^2 [u_{ij} + k - 1] 
\label{first}
\end{multline}

\begin{multline}
\pl_{1\leq j<i\leq kL}[v_{ij}]\rightarrow \\
\pl_{1\leq j<i\leq L}
[v_{ij}-k+1][v_{ij}-k+2]^{2}\cdots
[v_{ij}]^{k} \cdots 
[v_{ij}+k-2]^{2}[v_{ij}+k-1] \\
\times 
{\ll \pl_{p=1}^{k-1}[-p]_p \rr}^{L} 
\label{second}
\end{multline}

\noindent where the last factor in equation \ref{second} comes from 
diagonal terms such as $[-x_{2}+x_{1}]\rightarrow [-(x_{1}+1)+x_{1}]=[-1]$. 
Using equations \ref{first} and \ref{second}, the RHS of equation 
\ref{starting-point} becomes 

\begin{equation}
\frac{
\pl_{1 \leq i, j \leq L}
{\ll \pl_{p=1}^{k}   [-x_i + y_j + p]_{k+1} \rr}
{\ll \pl_{p=0}^{k-1} [-x_i + y_j + p]_{k-1} \rr}
det \ll M^{k/2\times k/2}_{kL\times kL} \rr
}
{
{\ll
\pl_{p=1}^{k-1}
[-p]_p
\rr}^{2L}
\ll \pl_{1 \leq j<i \leq L} \pl_{p=0}^{k-1} [-x_{i}+ x_{j}+p]_{k} \rr
\ll \pl_{1 \leq i<j \leq L} \pl_{p=0}^{k-1} [-y_{i}+ y_{j}+p]_{k} \rr
}
\end{equation}

To normalize, we divide by a factor so that the weight of the 
$c+$ vertex in the spin-$k/2$ model is $[k]_k$, and multiply 
by a factor that accounts for the change in symmetry structure 
of the determinant. Together, these are 

$$
\frac{
\ll
\pl_{p=1}^{k-1}
[-p]_p
\rr^{2L}
}
{
\ll  
\pl_{1 \leq i,j \leq L} 
\pl_{p=0}^{k-1} 
[-x_i+ y_j+p]_{k-1}
\rr
} 
$$

\noindent and we end up with the general result in equation 
\ref{main-result}. 

\subsection*{Acknowledgements} 

We thank A Dow and T Welsh for reading the manuscript and for 
helping us to improve the presentation. AC was supported by an 
Australian Postgraduate Award (APA), OF by the Australian Research 
Council (ARC), and NK by grant \# ANR-05-BLAN-0029-01.   



\begin{thebibliography}{99}

\bibitem{korepin}
{\sc V~E~Korepin},
{\em Commun. Math. Phys.} {\bf 86} (1982) 391--418.

\bibitem{izergin}
{\sc A~G~Izergin},
{\em Sov. Phys. Dokl.} {\bf 32} (1987) 878--879. 

\bibitem{slavnov}
{\sc N~Slavnov}, 
{\em Theor. Math. Phys.} {\bf 79} (1989) 502--8. 

\bibitem{rks}
{\sc  N~Reshetikhin, Kulish and Sklyanin},
{\em Lett. Math. Phys.} {\bf 5}(1981) 3993--403.

\bibitem{djkmo}
{\sc  A~Date, M~Jimbo, A~Kuniba, T~Miwa and M~Okado},
{\em Advanced Studies in Pure Mathematics} {\bf 16} (1988) 17--122.

\bibitem{korepin-book} 
{\sc V~E~Korepin, N~M~Bogoliubov, and A~G~Izergin}, 
{\em Quantum Inverse Scattering Method and Correlation Functions}, 
Cambridge University Press, 1993.

\bibitem{bressoud-book} 
{\sc D~M~Bressoud},
{\em Proofs and Confirmations: 
The story of the Alternating Sign Matrix Conjecture},
Cambridge University Press, 1999.

\bibitem{baxter-book}
{\sc R~J~Baxter},
{\em Exactly solved models in statistical mechanics},
Academic Press, 1982.

\bibitem{fz}
{\sc A~B~Zamolodchikov and V~A~Fateev},
{\em Soviet J Nuclear Phys}, {\bf 32} (1980) 581--590.

\bibitem{lima-santos}
{\sc A~Lima-Santos}, 
{\em J. Phys. {\bf A32}} (1999) 1819--48. 


\bibitem{fkow}
{\sc O~Foda, N~Kitanine, W~Orrick and T~A~Welsh},
{\em unpublished}.

\end{thebibliography}
\end{document}